\documentclass[12pt,a4paper]{article}

\usepackage{amsmath,amsthm,amssymb,amscd,a4wide}

\usepackage{dsfont}
\usepackage{mathrsfs}
\usepackage{setspace}
\usepackage{hyperref}

\newcommand{\ud}{\mathrm{d}}

\numberwithin{equation}{section}

\newtheorem{theorem}{Theorem}[section]

\newtheorem{prop}[theorem]{Proposition}

\newtheorem{remark}[theorem]{Remark}

\theoremstyle{definition}

\numberwithin{equation}{section}

\begin{document}

\thispagestyle{empty}

\vspace*{1cm}

\begin{center}

{\LARGE\bf On a non-separable quantum many-particle system on the half-line} \\

\vspace*{2cm}

{\large Joachim Kerner \footnote{E-mail address: {\tt Joachim.Kerner@fernuni-hagen.de}} and Tobias M\"{u}hlenbruch \footnote{E-mail address: {\tt Tobias.Muehlenbruch@fernuni-hagen.de}} }%

\vspace*{5mm}

Department of Mathematics and Computer Science\\
FernUniversit\"{a}t in Hagen\\
58084 Hagen\\
Germany\\

\end{center}

\vfill

\begin{abstract} In this paper we will report on a one-dimensional, non-separable quantum many-particle system introduced in \cite{1,12}. It consists of two (distinguishable) particles moving on the half-line $\mathbb{R}_+$ being subjected to two different kinds of two-particle interactions: singular  many-particle interactions localised at the origin and a binding-potential leading to a molecular-like state. We will formulate the model precisely, obtaining a well-defined self-adjoint operator (the Hamiltonian of our system) and elaborate on its spectral properties. In addition, we will present possible directions for future research.
\end{abstract}

\newpage

	\section{Introduction}
	
	The origin of the model presented in this paper is found in an investigation of many-particle quantum chaos using (compact) quantum graphs \cite{5,6}. Since the motion of a single particle on a quantum graph with complex topology becomes chaotic through the scattering of the particle in the vertices of the graph \cite{4}, the idea was to introduce many-particle interactions which are present only whenever at least on particle hits a vertex. In this way the scattering properties and hence the dynamics of the system are affected in a way to encode many-particle correlations which then enables one to study many-particle quantum chaos.
	
	The one-particle configuration space of the model presented in this paper is the half-line $\mathbb{R}_+=[0,\infty)$ and hence a simple, non-compact quantum graph (as we will see later, it is the non-compactness of the graph which generates a rich spectral theory). Furthermore, the Hamiltonian can be formally written as
	\begin{equation}\label{Hamiltonian}
	H=-\frac{\partial^2}{\partial x^2}-\frac{\partial^2}{\partial y^2}+v(x,y)\left[\delta(x)+\delta(y) \right]+v_{b}(|x-y|)\ ,
	\end{equation}
	$v:\mathbb{R}^2 \rightarrow \mathbb{R}$ being symmetric and $v_{b}:\mathbb{R}_+ \rightarrow [0,\infty]$ being the binding-potential leading to a molecular-like state. We note that the original model as presented in \cite{1} was formulated without binding-potential. It was later added in \cite{12}, connecting the model to the one discussed in \cite{2}. 
	
	The nature of the singular two-particle interactions considered can be directly understood from the formal expression \eqref{Hamiltonian} and the properties of the $\delta$-distribution. Indeed, since the $\delta$-distribution is supported on the origin only, these two-particle interactions are present only whenever at least one of the particles is situated at the origin. Quite surprisingly, although one is tempted to regard such singular interactions as a small perturbation only, we will see that this is not always the case. Indeed, as a main result, it was shown in \cite{12} that the singular interactions (given they are repulsive and strong enough) lead to a destruction of the ground state of the molecule and hence destabilise the system.
	
	Finally, we would like to mention that the model presented in this paper has various possible applications in different areas of theoretical and applied physics. On the one hand, it might serve as a toy-model to describe electrons moving in composite wires \cite{7} while, on the other hand, it is of theoretical interest by being an approachable many-particle model. In addition, it provides an example of a non-separable quantum many-particle model which are, despite their importance, otherwise only rarely discussed \cite{8,9}. 
	\section{The model}\label{Sec1}
	As in \cite{12} the binding-potential $v_b$ is chosen to be
	\begin{equation}\label{BindingPotential}
	v_b(|x-y|):=
	\begin{cases}
	0 \quad \text{if} \quad |x-y| \leq d\ , \\
	+\infty \quad \text{else}\ ,
	\end{cases}
	\end{equation}
	$0 < d \leq \infty$ characterising the ``size'' of the molecule. Due to the presence of the binding-potential, the two-particle configuration space is reduced from $\mathbb{R}^2_+$ to $\Omega$ which is given by
	\begin{equation}
	\Omega:=\{(x,y) \in \mathbb{R}^2_+\ | \ |x-y| \leq d\}\ .
	\end{equation}
	For later purposes we also define
	\begin{equation}
	\partial \Omega_{\sigma}:=\{(x,y) \in \Omega\ |\  x=0 \quad \lor \quad y=0 \}\ ,
	\end{equation}
	and
	\begin{equation}
	\partial \Omega_{D}:=\{(x,y) \in \Omega\ |\  |x-y|=d \}\ ,
	\end{equation}
	so that the boundary $\partial \Omega$ decomposes into $\partial \Omega_{\sigma}$ and $\partial \Omega_{D}$. Now, in order to arrive at a rigorous realisation of \eqref{Hamiltonian} we construct a suitable quadratic form on the Hilbert space $L^2(\Omega)$. We set
	\begin{equation}\label{QuadraticForm}
	q_{d}[\varphi]:=\int_{\Omega}|\nabla \varphi|^2 \ \mathrm{d}x -\int_{\partial \Omega_{\sigma}} \sigma(y)|\varphi_{bv}|^2 \ \mathrm{d}y\ ,
	\end{equation}
	being defined on the domain $\mathcal{D}_q:=\{ \varphi \in H^1(\Omega): \varphi|_{\partial \Omega_{D}}=0 \}$. We note that the Dirichlet boundary conditions along $\partial \Omega_{D}$ are induced by \eqref{BindingPotential}. Furthermore, we set $\sigma(y):=-v(0,y)$ and $\varphi_{bv}=:\varphi|_{\partial \Omega_{\sigma}}$ which are well defined according to the trace theorem for Sobolev functions \cite{11}.
	\begin{remark}\label{RemarkBoundaryConditions}
		The quadratic form \eqref{QuadraticForm} corresponds to a variational formulation of a bound\-ary-value problem for the two-dimensional Laplacian $-\Delta$ on $\Omega$
		with coordinate dependent Robin boundary conditions along $\partial \Omega_{\sigma}$ and Dirichlet boundary conditions along $\partial \Omega_{D}$. Indeed, the boundary conditions along $\partial \Omega_{\sigma}$ read
		\begin{equation}\label{RBConditions}\begin{split}
		\frac{\partial \varphi}{\partial n}(0,y)+\sigma(y)\varphi(0,y)&=0\ , \quad \text{and} \\
		\frac{\partial \varphi}{\partial n}(y,0)+\sigma(y)\varphi(y,0)&=0\ ,
		\end{split}
		\end{equation}
		for a.e. $y \in [0,d]$. Here $\frac{\partial}{\partial n}$ denotes the inward normal derivative along $\partial \Omega_{\sigma}$.

	\end{remark}
	The following was proved in \cite{1,12}.
	\begin{theorem} Let $\sigma \in L^{\infty}(0,d)$ with $0 < d \leq \infty$ be given. Then $q_{d}[\cdot]$ is densely defined, closed and semi-bounded from below.
	\end{theorem}
	Hence, according to the representation theorem for quadratic forms \cite{10}, there exists a unique self-adjoint operator being associated with $q_{d}[\cdot]$. This operator, being the Hamiltonian of our systems, shall be denoted by $-\Delta^d_{\sigma}$ and his domain by $\mathcal{D}(-\Delta^d_{\sigma}) \subset \mathcal{D}_q$.
	\begin{remark} We note that the case $d=\infty$ corresponds to the model where no binding pontential is considered in \eqref{Hamiltonian}. In other words, the operator $-\Delta^\infty_{\sigma}$ is the self-adjoint Hamiltonian of the model discussed in \cite{1}.
	\end{remark}
	\section{Spectral properties of the operator $-\Delta^d_{\sigma}$}
	As mentioned above, due to the non-compactness of the half-line, the spectrum $\sigma(-\Delta^d_{\sigma})$ cannot expected to be purely discrete. Indeed, using the method of Weyl sequences for forms \cite{14}, the following was established in \cite{1,12}.
	\begin{theorem}\label{Essential} Let $\sigma \in L^{\infty}(0,d)$ with $0 < d < \infty$ be given. Then 
		\begin{equation}
		\sigma_{ess}(-\Delta^d_{\sigma})=[\pi^2/2d^2,\infty)\ .
		\end{equation}
		Furthermore, if $d=\infty$ and $\sigma \in L^{\infty}(\mathbb{R}_+)$ is such that $\lim_{y \rightarrow \infty}|\sigma(y)|=0$, then 
		\begin{equation}
		\sigma_{ess}(-\Delta^d_{\sigma})=[0,\infty)\ .
		\end{equation}
	\end{theorem}
	In the case of vanishing binding-potential, i.e., $d=\infty$, Theorem~\ref{Essential} shows that the essential spectrum starts at zero given the modulus of the boundary potential $\sigma$ decays to zero as $y \rightarrow \infty$. In particular, this holds independently of the sign of $\sigma$, i.e., the singular interactions could be attractive ($\sigma > 0$) or repulsive ($\sigma <0$) or anything in between. A natural question is then the following: Given the singular interactions are ``attractive enough'', can eigenvalues below the bottom of the essential spectrum be expected? And what about the ground state energy of the system?
	\begin{theorem}\label{DiscreteSingular}\cite{1} For $d=\infty$ let $\sigma \in L^1(\mathbb{R}_+)$ be such that $\inf \sigma_{ess}(-\Delta^\infty_\sigma)=0$ and 
		\begin{equation}
		\int_0^\infty \sigma(y)\ \ud y > 0\ .
		\end{equation}
		Then there exists an eigenvalue at negative energy. 
		
		Furthermore, let $E_\sigma:=\inf \sigma(-\Delta^\infty_\sigma)$ denote the ground state energy. Then
		\begin{equation}
		-2\|\sigma\|^2_{\infty}\leq E_\sigma \leq -2\|\sigma\|^2_{\infty}+8\|\sigma\|^2_{\infty}\int_0^\infty \left[\|\sigma\|_{\infty}-\sigma(y)  \right]e^{-2\|\sigma\|_{\infty}y} \ \ud y\ .
		\end{equation}
	\end{theorem}
	We now turn to the case of non-vanishing binding-potential, i.e., $0 < d<\infty$. In the case of vanishing singular interactions ($\sigma \equiv 0$), $-\Delta^d_\sigma$ is a positive operator and therefore cannot possess negative spectral values. However, according to Theorem~\ref{Essential}, the essential spectrum starts only at $\pi^2/2d^2$ and for this reason one might ask the following question: do there exist non-negative eigenvalues with value smaller than $\pi^2/2d^2$? Using the results of \cite{3}, the following could be established in \cite{12}.
	\begin{theorem}\label{EigenvalueNoSingularBound} Assume that $0 < d < \infty$ and $\sigma\equiv 0$. Then 
		\begin{equation}
		\sigma_d(-\Delta^d_\sigma)\neq \emptyset\ .
		\end{equation}
		In other words, there exists a non-negative eigenvalue below the bottom of the essential spectrum.
	\end{theorem}
	\begin{remark} Theorem~\ref{EigenvalueNoSingularBound} is quite remarkable from a physical point of view since the molecule can remain localised around the origin without any attractive interactions present. In other words, the existence of such a bound states is a purely quantum mechanical effect which is induced by the geometry of the one-particle configuration space. If the one-particle configuration space was the whole line instead of the half-line, no such bound state would exist.
	\end{remark}
	We now investigate the stability of the discrete part of the spectrum in the presence of singular two-particle interactions. From a physical point of view, one expects the discrete spectrum to remain non-empty given the singular two-particle interactions are purely attractive or at least not too strong if repulsive. 
	\begin{prop}\label{Corollary}\cite{12} For given $\sigma\in L^{\infty}(0,d)$ and $0  <d < \infty$ the following holds: 
		\begin{enumerate}
			\item[1.] If $\sigma(y) \geq 0$ for almost every $y\in [0,d]$, then $\sigma_d(-\Delta^d_{\sigma}) \neq \emptyset$\ .
			\item[2.] There exists a constant $c > 0$ such that $\sigma_d(-\Delta^d_{\sigma}) \neq \emptyset$ for all $\sigma$ with $\|\sigma\|_{\infty} < c$.
		\end{enumerate}
	\end{prop}
	Finally, as a main result, the following was proved in \cite{12}.
	\begin{theorem}\label{TheoremDiscreteSpectrumRepulsive} Let $0 <d < \infty $ be given. Then there exists a constant $\gamma < 0$ such that $\sigma_d(-\Delta^d_{\sigma})=\emptyset$ for all $\sigma \in L^{\infty}(0,d)$ for which $\sigma(y) \leq \gamma$ holds for almost every $y \in [0,d]$. 
	\end{theorem}
	As already referred to in the introduction, Theorem~\ref{TheoremDiscreteSpectrumRepulsive} shows that the discrete part of the spectrum becomes trivial given the singular two-particle interactions are repulsive and strong enough. In this case, the singular two-particle interactions destabilise the system by leading to a destruction of the (isolated) ground state of the molecule and hence cannot be treated as a small perturbation only. 
	\section{Outlook}
	As mentioned earlier, the model presented is interesting from a theoretical point of view since it provides an example of an approachable many-particle system. For example, in \cite{16} the resolvent of the Hamiltonian \eqref{Hamiltonian} without binding-potential is constructed explicitly and the two-particle scattering properties are discussed in detail.
	
	It would be also be of interest to investigate the Cauchy problem for the Schr\"{o}dinger equation corresponding to a family of Hamiltonians $-\Delta^d_{\sigma(t)}$ (see, e.g., \cite{14}). What regularity of the map $t \mapsto \sigma(t)$ is needed in order to prove existence of a strong solution? How does the time-dependence of $\sigma$ affect the stability of the system?
	
	Finally, one could study the mean-field behaviour of the $N$-particle (mean-field) Hamiltonian being associated with \eqref{Hamiltonian} (see, e.g., \cite{15}). Is it possible to approximate the exact time-evolution of an $N$-particle state $\varphi_0\otimes ...\otimes \varphi_0$ with $\varphi_0 \in L^2(\mathbb{R}_+)$ by a tensor product $\varphi_t\otimes ...\otimes \varphi_t$, $\varphi_t$ being the solution to some non-linear Schr\"{o}dinger equation? Due to the singular nature of the two-particle interactions, is it possible to derive better convergence estimates?
	\section*{Acknowledgements}
	JK would like to thank S. Egger for helpful discussions.  We would also like to thank M.~L.~Glasser for his interest in our model and for providing stimulating references.

\end{document}